\documentclass[12pt]{article}

\normalbaselines

\catcode `\@=11
\@addtoreset{equation}{section}

\def\section{\@startsection {section}{1}{\z@}{-3.5ex plus -1ex minus 
-.2ex}{2.3ex plus .2ex}{\normalsize\bf}} 

\def\subsection{\@startsection{subsection}{2}{\z@}
    {-3.25ex plus -1ex minus -.2ex}{1.5ex plus .2ex}{\normalsize\bf}}
\def\subsubsection{\@startsection{subsubsection}{3}{\z@}{-3.25ex plus -1ex 
     minus -.2ex}{1.5ex plus .2ex}{\normalsize\sl}} 

\font\mybb=msbm10  at 12pt
\def\bb#1{\hbox{\mybb#1}}

\font\mybbi=msbm10  at 9pt
\def\bbi#1{\hbox{\mybbi#1}}
\def\Z{\bb Z}

\def\BC{\bb C}
\def\_\BC{\bbi C}
\def\RR{\bb R}
\def\Q{\bb Q}
\newcommand{\om}{\omega}
\newcommand{\Om}{\Omega}

\newcommand{\ot}{\otimes}
\newcommand{\De}{\Delta}

\newcommand{\bra}{\begin{array}}
\newcommand{\era}{\end{array}}
\newcommand{\be}{\beta}

\newcommand{\te}{\theta}
\newcommand{\beq}{\begin{equation}}
\newcommand{\eeq}{\end{equation}}
\newcommand{\bqn}{\begin{eqnarray}}
\newcommand{\eqn}{\end{eqnarray}}

\newcommand{\pa}{\partial}
\newcommand{\al}{\alpha}

\newcommand{\ka}{\kappa}

\title{\Large 
       \bf A Coherent-State Approach to Two-dimensional Electron Magnetism }

\author{\bf {J. P. Gazeau$^{\S}$}\thanks{e-mail: gazeau@ccr.jussieu.fr},
            {P. Y.  Hsiao$^{\S}$}\thanks{e-mail: hsiao@ccr.jussieu.fr}, 
        and {A. Jellal$^{\P}$}\thanks{e-mail: jellal@(ictp.trieste.it, youpy.co.uk)}}

\begin{document}
\maketitle
\begin{center}
{\it  $^{\S}$  Laboratoire de Physique Th\'eorique de la Mati\`ere 
              Condens\'ee, Boite 7020}\\  
{\it Universit\'e Paris 7 Denis Diderot},
{\it 2 Place Jussieu, 75251 Paris Cedex 05 France}\\  
{\it $^{\P}$ {\it Laboratory of High Energy Physics, Faculty of Sciences}\\
{\it Ibn Battouta Street, P.O. Box 1014} \it{Rabat-Morocco}}
\end{center}

\begin{abstract}
We study in this paper the possible occurrence of orbital magnetim for
two-dimensional electrons confined by a harmonic potential in various regimes
of temperature and magnetic field.  
Standard coherent state families are used for calculating symbols of various 
involved observables like thermodynamical potential, magnetic moment, or spatial
distribution of current. 
Their expressions are given in a closed form and the resulting Berezin-Lieb 
inequalities provide a straightforward way to study magnetism in various limit
regimes. 
In particular, we predict a paramagnetic behaviour in the thermodynamical limit
as well as in the quasiclassical limit under a weak field. 
Eventually, we obtain an exact expression for the magnetic moment which yields
a full description of the phase diagram of the magnetization. 
\end{abstract}


\section{Introduction}
In a recent paper, Ishikawa and Fukuyama \cite{ifu} describe the possible 
orbital magnetism for two-dimensional electrons confined by a harmonic 
potential in various regimes of temperature and magnetic field. 
They afford a quite large complement of information in regard to the previous
paper \cite{yofu} devoted to the same subject. 
According to the range of values assumed by the relative ratios between the
three characteristic energy scales present in the model, namely the 
thermodynamical unit $k_B T$, the magnetic quantum $\hbar \om_c$, and the 
harmonic quantum $\hbar \om_0$, they explain the existence of the 
different magnetic regimes.
As a matter of fact, they distinguish between the ``Mesoscopic Fluctuation'' 
regime 
($k_B T \stackrel{<}{\sim}\hbar (\sqrt{\om_c^2 + 4\om_0^2} - \om_c)/2$), 
the ``Landau Diamagnetism'' regime 
($k_B T \stackrel{>}{\sim}\hbar(\sqrt{\om_c^2 + 4\om_0^2} + \om_c)/2$) , 
and the ``de Haas-van Alphen'' regime. 
Their studies rest upon the derivation of an approximate closed formula for the
thermodynamical potential $\Om$
and from which they are able to get the magnetic moment.
The isotropic two-dimensional harmonic potential 
offers us  a complete analytical treatment of the problem
whereas the orbital magnetism is expected not to depend strongly 
on the shape of the confinement. 
The crucial point in our derivation is that the Fermi-Dirac function is a
fixed point of the Fourier transform.
Exact series expansions ensue by  simple application of the residue 
theorem as shown in the appendix.  
Our results for the thermodynamical potential, the orbital magnetic moment,
and the spatial distibution of currents are easily tractable and analyzable
in comparison with those given in Ref.\cite{ifu}. 

One can show that the Hamiltonian under a perpendicular magnetic field
can be decomposed into the sum of two harmonic oscillators with 
 frequencies $\om_+$ and $\om_-$ respectively. 
This fact makes the present model completely integrable and also makes  
the use of the coherent states (CS)\cite{klaska} possible. 
These coherent states are defined as  tensor products of  two
standard one-dimensional coherent states and read as: 
$\{\mid z_d,z_g\,\rangle=\,\mid z_d\,\rangle\,\ot\mid z_g\, \rangle,\ z_d,z_g
\in \BC \}$. 
This family of states provides us with a resolution of the unity and
in this way we are allowed to use the so-called  symbol calculus
{\it \`a la} Berezin-Lieb-Perelomov \footnote{ 
See \cite{ber} and Lieb in \cite{fks}. 
For a recent up-to-dated review on  CS, specially on their group representation aspects 
we also refer to \cite{aag}. 
Notice that here the operator calculus in Hilbert
space is replaced by the functional calculus involving CS parameters.}.
Consequently, the Berezin-Lieb inequalities give us the opportunity to explore the behavior of
the magnetization of our model in  extreme conditions 
like the thermodynamical limit or the quasi-classical limit. 

In the next section, we shall give a short review about the physical model. 
The eigenvalues and eigenvectors of the Hamiltonian will be obtained by 
algebraic and analytical methods respectively.
In particular, we shall explain the symmetries of the problem and their 
relevance in various approximations of the model.  
In Section III, the construction of the coherent states is presented 
and the main CS algebraic and functional properties are briefly recalled there. 
Section IV is devoted to the thermodynamical potential $\Om$. 
First we apply the Berezin-Lieb inequalities and  study the magnetic moment 
in the different limit regimes.  
We then give the exact expression of $\Om$. 
The core of the paper lies in Section V, in which we establish exact
expressions  for the magnetic moment and the average number of electrons 
and discuss their behavior in different temperature and magnetic field regions. 
We show in Section VI how coherent symbol calculus yields a Fourier integral
expression for the radial distribution of the current. 
This integral can be also given as an expansion series  with the 
aid of the residue theorem.  
In the conclusion we shall give some remarks and comments on the possible 
extension of our approach to the other systems of physical interest.   

\section{Hamiltonian and Symmetries} 
The Hamiltonian for two-dimensional spinless electrons confined by an isotropic
harmonic potential and submitted to a constant perpendicular magnetic field is
 written as
\beq
{\cal H}= \frac{1}{2m}({\bf P}+\frac{e}{c}{\bf A})^2
         +\frac{1}{2} m\om_0^2{\bf R}^2,
\eeq
where Coulomb interactions are neglected. 
This model is called the Fock-Darwin Hamiltonian in the litterature \cite{foda}.
The radius of the system, $R_m$, is classically defined as 
\beq
\frac{1}{2}m\om_0^2 R_m^2 = \mu,
\eeq
where $\mu$ is the chemical potential, and we shall give in Section IV a full
quantum statistical mechanics interpretation of this relation.
We shall work with the symmetric gauge 
${\bf A}= \frac{1}{2}{\bf H}\times {\bf R} $.
We first solve the eigenvalue and eigenvector problems by using the underlying 
Weyl-Heisenberg symmetries, and we shall next consider eigenvectors in another 
system of coordinates by following a more classical analytical method.

\subsection{Solutions through Weyl-Heisenberg symmetries} 
The algebraic structure is easily displayed if we adopt the method of 
separation of Cartesian variables. 
It leads to the following form of the Hamiltonian:
\beq
{\cal H} =  (\frac{P_x^2}{2m} + \frac{1}{8} m \om^2 X^2) 
          + (\frac{P_y^2}{2m} + \frac{1}{8} m \om^2 Y^2) 
          + \frac{\om_c}{2} L_z 
    \equiv {\cal H}_0 + \frac{\om_c}{2} L_z,
\eeq
where $\om_c = eB/mc$ is the cyclotron frequency, 
$\om = \sqrt{\om_c^2 + 4\om_0^2}$, and $L_z = XP_y - YP_x$. 
Here, we clearly see the splitting of the Hamiltonian into two
independent harmonic oscillator Hamiltonians plus the angular momentum operator.
Instead of directly using the oscillator annihilation operators:
\beq
a_x =\frac{1}{\sqrt{2}}(\frac{X}{l_0} + \frac{il_0}{\hbar} P_x), \ \ 
a_y =\frac{1}{\sqrt{2}}(\frac{Y}{l_0} + \frac{il_0}{\hbar} P_y), 
\eeq
we work with two new ones, which are linear superposition of $a_x$ and $a_y$:
\beq
a_d = \frac{1}{\sqrt{2}} (a_x - i a_y), \ \ 
a_g = \frac{1}{\sqrt{2}} (a_x + i a_y),  
\eeq 
where  $l_0 = \sqrt{2\hbar/m\om }$. 
Note that $a_d$ and $a_g$ are bosonic operators:  
$\lbrack a_d,a_d^{\dagger}\rbrack=\mbox{I}=\lbrack a_g,a_g^{\dagger}\rbrack$, 
and one has the useful identities:
\bqn
\nonumber 
X = \frac{l_0}{2}  ( a_d + a_d^{\dagger} + a_g + a_g^{\dagger}), \  
Y = \frac{l_0}{2i} (-a_d + a_d^{\dagger} + a_g - a_g^{\dagger}), \\ 
P_x = \frac{ \hbar}{2il_0} (a_d - a_d^{\dagger} + a_g - a_g^{\dagger}),\  
P_y = \frac{ \hbar}{2 l_0} (a_d + a_d^{\dagger} - a_g - a_g^{\dagger}).
\eqn
The operators ${\cal H}_0$ and $L_z$ then can be simply expressed in terms of 
the number operators $N_d = a_d^{\dagger} a_d$ and $N_g = a_g^{\dagger} a_g$ as:
\beq 
{\cal H}_0 = \frac{ \hbar \om}{2} (N_d + N_g +1), \ \ 
L_z = \hbar (N_d -N_g),
\eeq
and so
\beq 
{\cal H} =\frac{\hbar \om}{2}(N_d + N_g +1)+\frac{\hbar \om_c}{2} (N_d - N_g) 
         =\hbar(\om_+ N_d  + \om_-  N_g + \frac{\om}{2}), \label{2.8}
\eeq
where $\om_{\pm} = (\om \pm \om_c)/2$.
Eigenvalues  are trivially found from the  expression:
\beq 
{\cal H}\mid n_d, n_g \rangle = E_{n_d n_g} \mid n_d, n_g \rangle, \label{2.9}
\eeq
with $E_{n_d n_g} = \hbar(\om_+ n_d  + \om_-  n_g + \om/2)$ and where $n_d$ and $n_g$ 
are non-negative integers.
The corresponding eigenvectors are tensor products of single Fock oscillator states:
\beq 
\mid n_d, n_g \rangle = \mid n_d \, \rangle \, \ot \mid  n_g \,  \rangle =
\frac{1}{\sqrt{n_d!n_g!}}(a_d^{\dagger})^{n_d}(a_g^{\dagger})^{n_g} 
\mid 0, 0 \rangle.
\eeq

\subsection{$\mathbf{su}(2)$ and $\mathbf{su}(1,1)$ symmetries}
Two distinct dynamical symmetries exist on the level of quadratic observables. 
\begin{enumerate}
\item $ \mathbf{su}(2)$  {\bf symmetry }\\
The first one is of the type $\mathbf{su}(2)$ and is put into evidence by introducing 
the operators
\beq
S_+ = a_d^{\dagger} a_g, \ \  
S_- = a_g^{\dagger} a_d, \ \  
S_z = \frac{N_d - N_g}{2} = \frac{L_z}{2\hbar}.
\eeq
The commutation relations read as:
\beq
\lbrack S_+,S_- \rbrack = 2S_z, \ \  
\lbrack S_z,S_{\pm} \rbrack = \pm S_{\pm}, 
\eeq
and the invariant Casimir operator is given by
\beq
{\cal C} = \frac{1}{2}(S_+ S_- + S_- S_+) + S_z^2 
         = (\frac{N_d + N_g}{2})(\frac{N_d + N_g}{2} + 1). 
\eeq
Therefore, for a fixed value $\lambda = (n_d+n_g)/2$ of the operator 
$(N_d + N_g)/2 = {\cal H}_0/(\hbar \om) - 1/2$,
there exists a $(2\lambda + 1)$-dimensional UIR of $\mathbf{su}(2)$ in which the
operator $ S_z$ has its spectral values in the range 
$ -\lambda \leq \varsigma = (n_d - n_g)/2 \leq \lambda$.

\item  $ \mathbf{su}(1,1)$ {\bf symmetry }\\
The second one is of the type $\mathbf{su}(1,1)$:
\beq
T_+ = a_d^{\dagger} a_g^{\dagger}, \ \ 
T_- = a_g a_d, \ \
T_0 = \frac{1}{2}(N_d + N_g + 1)= \frac{{\cal H}_0}{\hbar \om},
\eeq
with
\beq
\lbrack T_+,T_- \rbrack = -2T_0, \ \
\lbrack T_0,T_{\pm} \rbrack = \pm T_{\pm}. 
\eeq
The Casimir operator then reads as:
\beq
{\cal D}=\frac{1}{2}(T_+ T_- + T_- T_+) - T_0^2 
        =-(\frac{N_d-N_g}{2}+\frac{1}{2})(\frac{N_d-N_g}{2}-\frac{1}{2})
        =-\frac{1}{4}(\frac{L_z}{\hbar}^2 -1). 
\eeq
When $n_d \geq n_g$,  
for a fixed value $\eta = (n_d-n_g + 1)/2 \geq 1/2 $ of the operator
$(N_d - N_g + 1)/2$, there exists a UIR of $\mathbf{su}(1,1)$ in the discrete series, in 
which the operator $ T_0 $ has its spectral values in the infinite range 
$ \eta, \eta +1, \eta + 2, \cdots$.
Alternatively, when  $n_d \leq n_g$, for a fixed value 
$\vartheta = (-n_d+n_g+1)/2 \geq 1/2$ of the operator $(-N_d+N_g+1)/2$, 
there also exists a UIR of $\mathbf{su}(1,1)$ in which the spectral value of the
operator $ T_0 $ runs in the infinite 
range $ \vartheta, \vartheta +1, \vartheta + 2, \cdots$.
\end{enumerate}

\subsection{Solutions through analytical derivation}
The analytical solutions are obtained by separation of polar coordinates in 
the stationary Schr\"{o}dinger equation 
\beq
{\cal H}\Psi(r,\te)
=[-\frac{\hbar^2}{2m}(\pa_r^2+\frac{1}{r}\pa_r +\frac{1}{r^2}\pa_{\te}^2)
  - i\frac{\hbar}{2}\om_c \pa_{\te} 
  + \frac{m}{8}\om^2 r^2            
 ]\Psi(r,\te) = E\Psi(r,\te).
\eeq
We consider $\Psi(r,\te)$ as an eigenfunction diagonal with respect to the 
conserved angular momentum and put $\Psi(r,\te)= R(r)e^{i\al\te}$. 
The function $R(r)$ is determined in terms of Laguerre polynomials. 
Explicitly we have 
\beq
\Psi(r,\te)=\Psi_{n,\al}(r,\te)
=(-1)^n \times \frac{1}{\sqrt{\pi} l_0} \sqrt{\frac{n!}{(n+\vert\al \vert)!}} 
\exp{\lbrack -\frac{r^2}{2l_0^2}\rbrack}\, 
(\frac{r}{l_0})^{\vert \al \vert}\, 
L_n^{(\vert \al \vert)}\lbrack\frac{r^2}{l_0^2} \rbrack\,
 e^{i\al\te}, \label{2.19}
\eeq
where $n=0, 1, 2,\cdots$ is the principal quantum number and
$\al=0, \pm 1, \pm 2 , \cdots$ the angular moment quantum number. 
The eigenenergies are obtained to be:
\beq
E_{n\al} = \hbar\om(n+\frac{\vert \al \vert+1}{2})+\frac{\hbar\om_c}{2}\al,
\eeq
Therefore, $n$ and $\al$ are related to $n_d$ and $n_g$ by the following 
relations:   
$$
n_d=n+ \frac{1}{2}(\vert \al \vert + \al), \ \mbox{and}\
n_g=n+ \frac{1}{2}(\vert \al \vert - \al).
$$
We shall denote the eigenfunction indifferently by 
$\Psi_{n,\al} (r,\te) =\langle r , \te \mid n, \al \rangle 
                      =\langle r , \te \mid n_d, n_g \rangle$.

\subsection{Filling the shells with fermions}
Each of the above two symmetries affords a way of ordering the pairs 
$(n_d, n_g)$.
However, they do not provide any hint for ordering the energy eigenvalues
$E_{n_d n_g}$ with the exception of two limiting cases: weak field one and 
strong field one.
\begin{enumerate}
\item {\bf Weak field case} \\
Suppose the average number of electrons $\langle N_e \rangle$ obeys 
$\om_c \langle N_e \rangle << \om_0$. 
In terms of the $\mathbf{su}(2)$ symmetry, the energy eigenvalues read as:
\beq
E_{n_d n_g}=\hbar\om\lambda+\hbar\om_c\varsigma+\frac{1}{2}\hbar\om,
\eeq
and can be approximated by
\beq
E_{n_d n_g} \approx \hbar \om_0 (2\lambda + 1) \equiv E_{\lambda}.
\eeq
Therefore, in the weak field limit, the $\mathbf{su}(2)$ symmetry becomes a true
symmetry of the Hamiltonian, which explains the degeneracy of order
$2\lambda + 1$ for the level $E_{\lambda}$.
Note that there are $(\lambda_0 + 1) (2\lambda_0 + 1)$ (spinless) electrons 
which fill the shells up to $\lambda_0$.
\item {\bf Strong field case}\\
In the limit of strong magnetic field $\om_c \gg \om_0$,
we have $E_{n_d n_g} \approx \hbar \om_c (n_d + \frac{1}{2})$. 
Therefore, for a given value of $n_d$, we have an infinite degeneracy labelled
by $n_g$ or by $\al = n_d - n_g \leq n_d$. 
The quantum number $n_d$ corresponds to the Landau level index
(as well as $n$ for negative $\al$).  
One can reinterpret it in terms of $\mathbf{su}(1,1)$ symmetry by noting that, 
for a given value of $\al \leq 0$,  
the energy eigenstates are ladder states for the discrete series 
representation labelled by $\vartheta = - \al/2 + 1/2$.
\item {\bf Generic intermediate case}\\
In the uncommensurate intermediate case, which means $\om_+/\om_- \notin \Q$
and no approximation is relevant,
we are faced to the problem of ordering the relatively dense (but not uniformly
discrete) set of eigenenergies:  
\beq
{\cal E}_{n_d n_g} \equiv \frac{E_{n_d n_g}}{\hbar\om_-}-\frac{\om}{2\om_-}
                     = \frac{\om_+}{\om_-}n_d + n_g.
\eeq
In the commensurate case, $\om_+/\om_- = p/q \in \Q$, degeneracy is possible:
\beq
E_{n_d n_g} = E_{n'_d n'_g} \ \mbox{iff} \ 
\frac{p}{q} = - \frac{n_g - n'_g}{n_d - n'_d}.
\eeq 
\end{enumerate}    

\section{Standard Coherent States}
\subsection{Definitions and Properties}
The fact that the eigenvectors issued from the algebraic method are just 
tensor products of Fock harmonic oscillator eigenstates allows one to construct
the corresponding coherent states in a standard way: 
\bqn
\nonumber 
\mid z_d,z_g\rangle\equiv\mid z_d\,\rangle\, \ot\mid z_g\,\rangle = 
\exp{\lbrack -\frac{1}{2}(\vert z_d \vert^2 + \vert z_g \vert^2)\rbrack }\, 
\sum_{n_d, n_g}\frac{z_d^{n_d}}{\sqrt{n_d!}}\frac{ z_g^{n_g}}{\sqrt{n_g!}}
\mid n_d, n_g \rangle \\
= \exp{\lbrack-\frac{1}{2}(\vert z_d \vert^2+\vert z_g\vert^2)\rbrack }\,
 e^{z_d a_d^{\dagger} + z_g a_g^{\dagger}} \mid 0, 0 \rangle .
\eqn 
These (normalized) coherent states obey some of the usual CS properties 
\cite{klaska}:

\begin{enumerate}
\item[P$_1$] (Eigenvector property)
$$
a_d \mid z_d, z_g \rangle = z_d \mid z_d, z_g \rangle,\ \ 
a_g \mid z_d, z_g \rangle = z_g \mid z_d, z_g \rangle.
$$
\item[P$_2$] (Minimal uncertainty relations)
$$
\De X \, \De P_x = \frac{\hbar}{2}, \  \ 
\De Y \, \De P_y = \frac{\hbar}{2},
$$
where $\De X\equiv\lbrack\langle z_d, z_g\mid X^2\mid z_d, z_g\rangle
 -(\langle z_d, z_g \mid X  \mid z_d, z_g \rangle)^2 \rbrack^{1/2}$, {\it etc}.
\item[P$_3$] (Temporal stability)
$$
e^{-i{\cal H}t/\hbar}\mid z_d, z_g\rangle 
=e^{-i\om t/2}\mid e^{-i(\om + \om_c)t/2}z_d,e^{-i(\om-\om_c)t/2}z_g\rangle.
$$
\item[P$_4$] (Action identity)
\beq
\check{{\cal H}}(z_d, z_g)
\equiv\langle z_d, z_g \mid {\cal H}\mid z_d, z_g \rangle 
=\hbar\lbrack\om_+\vert z_d\vert^2+\om_-\vert z_g\vert^2+\frac{\om}{2}\rbrack.
\label{3.2}
\eeq
The function $\check{{\cal H}}(z_d, z_g)$ has been called lower 
(resp. contravariant) symbol of the operator ${\cal H}$ by Lieb \cite{fks} 
(resp. by Berezin \cite{ber}). 
It will plays an important role in the present context.
\item[P$_5$] (Overlapping)
$$
\langle z'_d, z'_g\mid z_d,z_g\rangle 
= e^{i\Im (z_d \bar{z}'_d+ z_g \bar{z}'_g)}
  e^{-(\vert z_d - z'_d \vert^2 + \vert z_g - z'_g \vert^2)/2}.
$$
\item[P$_6$] (Resolution of the unity)
$$
\mbox{I}=\frac{1}{\pi^2}\int_{\BC^2}\mid z_d,z_g\rangle\langle z_d, z_g \mid\,
          d^2z_d \, d^2z_g.
$$
\end{enumerate}
The last property is also crucial in our context.
For any observable $A$ with suitable operator properties (traceclass, ...) 
there exists a unique upper (or covariant) symbol $\hat{A}(z_d, z_g)$ defined by
\beq
A=\frac{1}{\pi^2}\int_{\BC^2}\, \hat{A}(z_d, z_g)\, 
\mid z_d, z_g\rangle\langle z_d, z_g \mid\, d^2z_d \, d^2z_g.
\eeq
For instance, upper symbols for number operators are given by 
\beq
\hat{N_d}(z_d, z_g) = \vert z_d \vert^2 -1,\ \ 
\hat{N_g}(z_d, z_g) = \vert z_g \vert^2 -1, 
\eeq
and hence the upper symbol for our Hamiltonian (\ref{2.8})
\beq
\hat{{\cal H}}(z_d, z_g)=\hbar\lbrack\om_+\vert z_d \vert^2+\om_-\vert z_g 
\vert^2 -\frac{\om}{2} \rbrack.  \label{3.6}
\eeq
Finally, we should mention the useful trace identity for a traceclass
observable $A$:
\beq
\mbox{Tr}A =\frac{1}{\pi^2}\int_{\BC^2}\, \check{A}(z_d,z_g)\, d^2z_d\, d^2z_g 
=\frac{1}{\pi^2}\int_{\BC^2}\, \hat{A}(z_d, z_g)\, d^2z_d\, d^2z_g,
\eeq 
where $\check{A}(z_d, z_g)\equiv\langle z_d, z_g\mid  A\mid z_d, z_g\rangle$

\subsection{Fock-Bargman space}
The coherent state function $\langle r,\te\mid z_d, z_g\rangle$ is, up to an 
exponential factor, the integral kernel for the isometry mapping the Hilbertian
span $\mbox{L}^2(\RR^2)$ of the set of eigenfunctions (\ref{2.19}) onto the 
so-called Fock-Bargman space, {\it i.e.}  the Hilbert space 
\beq
{\cal F}{\cal B}\equiv\mbox{L}_{\scriptsize \mbox{entire}}^2
(\BC^2,\, 
\frac{1}{\pi^2}e^{-(\vert z_d\vert^2+\vert z_g\vert^2)}\, d^2z_d\, d^2z_g)
\eeq 
of entire two complex variable functions $f(z_d, z_g)$ that are square 
integrable with respect to the measure
$\pi^{-2}e^{-(\vert z_d \vert^2 + \vert z_g \vert^2)}\, d^2z_d\, d^2z_g$. 
This unitary mapping 
$\mbox{L}^2(\RR^2)\ni\Psi(r,\te)\rightarrow\Phi(z_d,z_g)\in {\cal F}{\cal B}$ 
and its reciprocal
are explicitely given by
\bqn
\Psi(r,\te) & = &
\frac{1}{\pi^2}\int_{\scriptsize\BC^2}{\cal K}(r,\te,\bar{z}_d,\bar{z}_g)
\Phi(z_d,z_g)e^{-(\vert z_d\vert^2+\vert z_g\vert^2)}\, d^2z_d\, d^2z_g
\label{3.8}\\
\Phi (z_d, z_g) &=& 
\int_{\scriptsize\RR^2}{\cal K}(r,-\te,z_d,z_g)\Psi(r,\te)r\,dr\,d\te.
\eqn
The kernel is given by the following generating function:
\bqn
\nonumber
{\cal K}(r,\te,z_d,z_g) &=& 
e^{(\vert z_d \vert^2+\vert z_g\vert^2)/2}\langle r,\te\mid z_d, z_g\rangle 
=\sum_{n_d,n_g}\frac{z_d^{n_d}}{\sqrt{n_d!}}\frac{ z_g^{n_g}}{\sqrt{n_g!}}
  \Psi_{n, \al} (r , \te) \\
&=& \frac{1}{\sqrt{\pi}l_0}e^{r^2/(2l_0^2)}\, 
e^{-(z_d -e^{-i\te}r/l_0)(z_g - e^{i\te}r/l_0)}.
\eqn
Note that normalized eigenkets (\ref{2.19}) are mapped to the corresponding 
normalized eigenstates 
\beq
\frac{z_d^{n_d}}{\sqrt{n_d!}}\frac{ z_g^{n_g}}{\sqrt{n_g!}} =
 e^{(\vert z_d\vert^2+\vert z_g\vert^2)/2}
\langle\bar{z}_d, \bar{z}_g \mid n_d, n_g \rangle  
\eeq
in the ${\cal F}{\cal B}$ representation.
Also note that in this representation the operators $a$ and $a^{\dagger}$ 
have the simple form:
\beq
a_d =\frac{\partial}{\partial z_d}, \ 
a^{\dagger}_g =\, (\mbox{multiplication by})\, z_d, \ 
a_g=\frac{\partial}{\partial z_g}, \ 
a^{\dagger}_g=z_g.
\eeq

\section{Thermodynamical potential}
Let us now enter the core of the physical question we addressed in the 
introduction. 
We assume that the total number $\langle N_e \rangle$ of electrons is large 
enough for making no appreciable difference between a grand canonical ensemble 
and a canonical one. 
Then the magnetic moment $M$ is given by
\beq
M = -\left(\frac{\pa \Om}{\pa H}\right)_{\mu}, \label{4.1}
\eeq
where $\Om$ is the thermodynamical potential,
\beq
\Om = -\frac{1}{\be}\mbox{Tr}\log{(1+e^{-\be({\cal H}-\mu)})} ,\label{4.2}
\eeq
with $\be = 1/(k_BT)$.
The average number of electrons is given by: 
\beq
\langle N_e\rangle=\sum_{n_d=0}^{\infty}\sum_{n_g=0}^{\infty} f(E_{n_d n_g}) 
= \mbox{Tr}f({\cal H}) = -\partial_{\mu} \Om,
\eeq
where $f(E) = 1/(1 + e^{\be (E - \mu)})$ is the Fermi distribution function.
The magnetic moment can be yielded from eqs.(\ref{4.1}) and (\ref{4.2}) and
reads as: 
\bqn
\nonumber 
\frac{M}{\mu_B} &=& -\frac{2}{\om}\,
\mbox{Tr}\frac{(N_d+1/2)\om_{+}-(N_g+1/2)\om_{-}}{(1+ e^{\be({\cal H}-\mu)})}\\ 
&=& -\frac{2}{\om}\, \sum_{n_d=0}^{\infty} \sum_{n_g=0}^{\infty}
\frac{(n_d+1/2)\om_{+}-(n_g+1/2)\om_{-}}
     {(1+\kappa_-^{-1}e^{\be\hbar(\om_+n_d+\om_-n_g)})},\label{4.4}
\eqn
where $\ka_{\pm} =\exp{(\be(\mu \pm \hbar \om/2))}=\ka_{\pm}(H,T)$,
and $\mu_B =\hbar e/(2mc)$ is the Bohr magneton. 
Despite its concise appearance, the computation of the double series (\ref{4.4})
is not easily tractable on a numerical level.
Hence we will give the preliminary estimates before presenting much more 
exploitable exact expressions.

\subsection{Berezin-Lieb inequalities for the thermodynamical potential}
First we note that $\log{(1 + e^{-\be({\cal H} - \mu)})}$ is a convex function 
of the positive Hamiltonian ${\cal H}$, and so we can apply the Berezin-Lieb 
inequalities to examine the quasi-classical behaviour of the thermodynamical 
potential.
These inequalities say that, for any convex function $g(A)$ of the observable 
$A$, we have  
\beq
\frac{1}{\pi^2}\int_{\BC^2}g(\check{A})\,d^2z_d\,d^2z_g 
\leq \mbox{Tr}g(A) \leq 
\frac{1}{\pi^2}\int_{\BC^2}g(\hat{A}) \, d^2z_d\,d^2z_g.
\eeq
Applying them to the (concave) thermodynamical potential leads to the 
inequalities:
\beq
-\frac{1}{\be\pi^2}\int_{\BC^2}\log{(1+e^{-\be(\hat{{\cal H}}-\mu)})}\,
d^2z_d \, d^2z_g \leq \Om \leq  
-\frac{1}{\be\pi^2}\int_{\BC^2}\log{(1+e^{-\be(\check{{\cal H}}-\mu)})}\, 
d^2z_d \, d^2z_g.
\eeq
Let us insert eqs.(\ref{3.2}) and (\ref{3.6}) and perform the angular 
integrations. It leads to the inequalities:
\bqn
\nonumber 
-\frac{1}{\be}\int_{0}^{\infty}du_d\,\int_{0}^{\infty} du_g\,
\log{(1+e^{-\be(\hbar(\om_+u_d+\om_-u_g-\frac{\om}{2})-\mu)})}\leq\Om,\\
\Om \leq -\frac{1}{\be}\int_{0}^{\infty}du_d\,\int_{0}^{\infty}du_g\, 
\log{(1+e^{-\be(\hbar(\om_+u_d +\om_- u_g+\frac{\om}{2})-\mu)})}, 
\eqn
where $u_d = \vert z_d \vert^2$ and $u_g = \vert z_g \vert^2$. 
Changing the integration variables, say 
$u= \be\hbar(\om_+ u_d +\om_- u_g),\, v=\be\hbar\om_+ u_d$, 
performing an integration by part, and introducing the control parameters 
$\ka_{\pm}$ defined in (\ref{4.4}), we easily reduce the above inequalities to 
the following ones:
\beq
\phi(\ka_+)\leq\Om\leq\phi(\ka_-). \label{4.8}
\eeq
where the function $\phi$ is given by:
\bqn
\nonumber 
\phi(\ka) &=& -\frac{\ka}{2\be(\be\hbar\om_0)^2}
\int_0^{\infty}\frac{u^2 e^{-u}}{1+\ka e^{-u}}\, du \\
&=& \left\{ \begin{array}{ll} 
 \frac{1}{\be(\be\hbar\om_0)^2}\,F_3(-\ka) & \mbox{for $\ka \leq 1$},\\
 \frac{1}{\be(\be \hbar \om_0)^2}
\left\lbrack -\frac{(\log{\ka})^3}{6}-\frac{\pi^2\log{\ka}}{6}+ F_3(-\ka^{-1})
\right\rbrack & \mbox{for $\ka >1$}.
 \end{array}
\right. \label{4.9}
\eqn
We have introduced here the function $F_s$, of the Riemann-Fermi-Dirac type, defined as: 
\beq
F_s(z)=\sum_{m=1}^{\infty} \frac{z^m}{m^s }.
\eeq
In the high temperature region $\vert \mu \pm \hbar \om/2 \vert \ll k_B T$ we have
$\ka_{\pm} \approx 1$. 
From (\ref{4.8}) we see that  the thermodynamical potential is approximately equal to:
$$  
\Om \approx k_BT\left(\frac{k_B T}{\hbar\om_0}\right)^2 F_3(-1)
\approx -0.901543\,k_BT\left(\frac{k_B T}{\hbar \om_0}\right)^2
$$
We now consider the more realistic case:
$\mu \gg \hbar \om/2$ and $\mu \gg k_B T$. 
Let us split the function $\phi$ into three parts: 
\beq
\phi(\ka_{\pm}) = A \mp \frac{\Delta}{2} + S_{\pm}, \label{4.11}
\eeq
with
\bqn
\nonumber 
A &=& 
-\frac{\mu}{2}\left\lbrack\frac{1}{3}\left(\frac{\mu}{\hbar \om_0}\right)^2 + 
\frac{1}{4}\left(\frac{\om}{\om_0}\right)^2+\frac{\pi^2}{3}
\left(\frac{k_B T}{\hbar \om_0}\right)^2\right\rbrack,\\
\nonumber 
\frac{\Delta}{2} &=& 
\frac{\hbar\om}{2}\left\lbrack\frac{1}{2}\left(\frac{\mu}{\hbar\om_0}\right)^2+ 
\frac{1}{24}\left(\frac{\om}{\om_0}\right)^2+\frac{\pi^2}{6}
\left(\frac{k_BT}{\hbar \om_0}\right)^2 \right\rbrack, \\ 
S_{\pm} &=& k_B T \left(\frac{k_B T}{\hbar \om_0}\right)^2  
F_3(-\exp{[-\be( \mu \pm \hbar \om /2)]}). 
\eqn
Then we see that $\Om$ lies in the interval 
$\lbrack A+S_+-\Delta/2,A+S_-+\Delta/2 \rbrack$. 
Replacing $S_{\pm}$ by the approximate expression 
\beq
S_0= k_B T \left( \frac{k_B T}{\hbar \om_0}\right)^2 F_3( - e^{-\be \mu}),
\eeq
and observing that  the ratio
\beq
\frac{\Delta}{\vert A + S_0 \vert} = \frac{\hbar \om}{\mu} \left\lbrack \frac{ 3 + \pi^2 \left(\frac{k_B T}{\mu}\right)^2 + \frac{1}{4} \left(\frac{\hbar
\om}{\mu}\right)^2}{ 1 + \pi^2 \left(\frac{k_B T}{\mu}\right)^2 + \frac{3}{4} \left(\frac{\hbar \om}{\mu }\right)^2 - \left(\frac{k_B T}{\mu}\right)^3 F_3( -
e^{-\be \mu}) } \right\rbrack
\eeq
tends to zero, we see that the thermodynamical potential can be estimated as:
\bqn
\nonumber
\Om & \approx & A + S_0   \\
&=&-\frac{\mu}{2}\left\lbrack\frac{1}{3}\left(\frac{\mu}{\hbar\om_0}\right)^2 
 +\frac{1}{4} \left(\frac{\om}{\om_0}\right)^2
 +\frac{\pi^2}{3}\left(\frac{k_BT}{\hbar \om_0}\right)^2\right\rbrack 
 +k_BT\left(\frac{k_BT}{\hbar\om_0}\right)^2 F_3(-e^{-\be\mu}).
\eqn
A similar asymptotic behaviour holds for the thermodynamical limit 
$\langle N_e \, \rangle \to \infty$. 
Indeed, in this quasiclassical regime, the average number of electrons is 
calculated to be: 
\bqn
\nonumber \langle N_e \, \rangle  &\approx& -\partial_{\mu} (A + S_0)\\
\nonumber & = & 
\left(\frac{\mu}{\hbar \om_0}\right)^2 \left\lbrack \frac{1}{2}   
+\frac{1}{8}\left(\frac{\hbar\om}{\mu}\right)^2 
+\frac{\pi^2}{6}\left(\frac{k_B T}{\mu}\right)^2
+\left(\frac{\om_0}{\om}\right)^2 \left(\frac{k_B T}{\mu}\right)^2 
F_2 (-e^{-\mu \be}) \right\rbrack\\
&\approx & \frac{1}{2} \left(\frac{\mu}{\hbar \om_0}\right)^2 \ 
\mbox{for $\mu \gg k_B T$ and $\mu \gg \hbar\om/2$}.
\eqn
The magnetic moment hence reads as: 
\beq
M = \chi_p H, \ \mbox{with} \ 
\chi_p = \mu \left(\frac{\mu_B}{\hbar \om_0}\right)^2. \label{4.17}
\eeq
Hence, we can assert that the system shows an orbital paramagnetism in this limit. At this 
point, we refer to  a recent work by Combescure and Robert \cite{coro} in which precise informations
are given  for the magnetisation of electron gas constrained by general confinement potentials.

\subsection{Exact expressions for the thermodynamical potential}
We now determine the thermodynamical potential in a precise way by applying 
the formulas (\ref{a.5}) and (\ref{a.7}) in the appendix. 
The function $\Theta(k)$ defined by (\ref{a.6}) takes the following  closed form:
\beq
\Theta(k)=\mbox{Tr}(e^{-(ik + 1)\frac{\be}{2}{\cal  H}}) 
=e^{-(ik+1)\frac{\be}{4}\hbar\om}\, 
\frac{1}{1 - e^{-(ik + 1)\frac{\be}{2}\hbar \om_+}}\, 
\frac{1}{1 - e^{-(ik + 1)\frac{\be}{2}\hbar \om_-}}.
\eeq
The Fourier integral representation for the thermodynamical potential hence
reads as:
\beq
\Om = -\frac{1}{\be }\int_{-\infty}^{+\infty} 
\frac{e^{-(ik+ 1)\frac{\be}{2}(\frac{\hbar\om}{2}-\mu)}}
     {2\cosh{\frac{\pi}{2}k}} \, 
\left(\frac{1}{ik+1}\right)
\left(\frac{1}{1-e^{-(ik+ 1)\frac{\be}{2}\hbar \om_+}}\right) 
\left(\frac{1}{1 - e^{-(ik + 1)\frac{\be}{2}\hbar \om_-}}\right) dk
\eeq
As indicated in the formula (\ref{a.6}), this Fourier integral is given 
as a series  by using the residue theorem. 
One can easily see  that the numbers
$(2m+1)i$, $m\in \Z$ are simple poles of $\mbox{sech}{\frac{\pi}{2}k}$,
$i$ is a double pole of $\Theta (k)$,
and $i+4\pi m/(\be\hbar\om_+)$, $i+4\pi m/(\be\hbar\om_-)$, $m\in \Z^{\ast}$
are simple or double poles of $\Theta(k)$ according to whether
$\om_+$ and $\om_-$ are uncommensurable or not (see Fig.1). 
In order to fulfill the requirements of the Jordan Lemma, 
one has to consider the following two cases: 
$\mu\leq\hbar\om/2$ and $\mu\geq\hbar\om/2$. 
In the first case we take an integration path lying in the lower half-plane and
involving only the simple poles $(2m+1)i$, $m<0$. 
It leads to the result: 
\beq
\Om = \frac{1}{4\be}\sum_{m=1}^{\infty} \frac{(-1)^m}{m}
\frac{e^{\beta \mu m}}
     {\sinh{(\frac{\be}{2}\hbar \om_+m)}\,\sinh{(\frac{\be}{2}\hbar \om_-m)}},
\label{4.20}
\eeq
which corresponds to the case $\ka\leq 1$ in Eq.(\ref{4.9}). 
In the second case, an integration path in the upper half-plane is chosen. 
It encircles all the other poles: $(2m+1)i$, $m\geq 0$, 
$i+4\pi m/(\be\hbar\om_+)$, $i+4\pi m/(\be\hbar\om_-)$, $m\in\Z^{\ast}$, 
as shown in Fig.1.
We present the result in a manner which will render apparent the various regimes:
\beq
\begin{array}{cccccccc}
\Om &=& &(\Om_L + \Om_{01})& + &\Om_{02}& + &\Om_{\mbox{\scriptsize osc}}\\
&=&2\pi i(&\overbrace{a_{-1}(i)}& +&\overbrace{\sum_{m\geq 1}a_{-1}((2m+1)i)}& 
 +&\overbrace{\sum_{m_{\pm}\not= 0}
 (a_{-1}(i+\frac{4\pi}{\be\hbar\om_{\pm}}m_{\pm})}).
\end{array}
\eeq
The first term is at the origin of the Landau diamagnetism:
\beq
\Om_L=\frac{\mu}{24}\left(\frac{\om_c}{\om_0}\right)^2 
     =-\frac{1}{2} \chi_L H^2,
\eeq
where $\chi_L=-\frac{1}{3}\mu\left(\frac{\mu_B}{\hbar\om_0}\right)^2 
\equiv -\frac{1}{3}D_0\mu_B^2$ is the Landau diamagnetic susceptibility. The coefficient
$D_0=\mu/(\hbar \om_0)^2$ can be interpreted as the density of states at 
Fermi energy. 
Notice that the value of $\chi_L$ is  equal to one third of the ``quasiclassical'' one
$\chi_p$ obtained in (\ref{4.17}), a feature which is reminiscent of which we encounter in the studies of 
free 3D electron gas. 
The second term, which gives no contribution to the magnetization, is written 
as:
\beq
\Om_{01}=-\frac{\mu}{6}\left\lbrack\left(\frac{\mu}{\hbar \om_0} \right)^2 
 +\pi^2\left(\frac{k_BT}{\hbar\om_0}\right)^2-\frac{1}{2}\right\rbrack.
\eeq 
The third term is given as:
\beq
\Om_{02}=\frac{1}{4\be}\sum_{m=1}^{\infty}\frac{(-1)^m}{m}
\frac{\exp{(- \frac{\mu}{k_B T} m)}}
{\sinh{(\frac{\hbar \om_+}{2k_B T}m)}\,\sinh{(\frac{\hbar \om_-}{2k_B T}m)}}.
\eeq
It becomes negligible at low temperature regime $k_BT\ll \mu$.
The sum of $\Om_L$ and $\Om_{01}$ is analogue to the term $A$ in Eq.(\ref{4.11})
and $\Om_{02}$ corresponds to $S_{\pm}$. 
The last term is responsible for the oscillatory behaviour. 
We need to distinguish between irrational values  of $\om_+/\om_-$ and 
rational ones:
\begin{itemize}
\item {\bf case} $\om_+/\om_- \not\in \Q$, 
\bqn
\nonumber \Om_{\mbox{\scriptsize osc}}&=& 
\frac{1}{2\be} \sum_{m=1}^{\infty} \frac{(-1)^m}{m} 
\left\lbrack\frac{\sin{(\frac{2 \mu}{\hbar \om_-}\pi m)}}
{\sin{(\frac{\om_+}{\om_-}\pi m)}\,
\sinh{(\frac{2k_B T}{\hbar\om_-}\pi^2 m)}} \right.\\
 &+ & \left. \frac{\sin{(\frac{ 2\mu}{\hbar \om_+}\pi m)}}
{\sin{(\frac{\om_-}{\om_+}\pi m)}\,
\sinh{(\frac{2 k_B T}{\hbar \om_+}\pi^2 m)}} \right\rbrack 
\equiv \Om_{\mbox{\scriptsize osc}}^- + \Om_{\mbox{\scriptsize osc}}^+.
\eqn
\item {\bf case} $\om_+/\om_-=p/q\in\Q,\ \gcd{(p,q)} = 1, \ 
\om_+/p = \om_-/q = 2l/(\hbar\be) \in\RR$, 
\bqn
\nonumber \Om_{\mbox{\scriptsize osc}} &=&  \frac{1}{2\be} \left\lbrack 
\sum_{m=1 ,\, m\not\equiv 0\, \mbox{\scriptsize mod}\, q}^{\infty} 
\frac{ (-1)^m}{m} \frac{\sin{(\frac{ 2\mu}{\hbar \om_-}\pi m)}}
{\sin{(\frac{\om_+}{\om_-}\pi m)}\,
\sinh{(\frac{2 k_B T}{\hbar \om_-}\pi^2 m)}} 
\right.\\
 &+ & \sum_{m=1 ,\, m\not\equiv 0 \, \mbox{\scriptsize mod}\, p}^{\infty} 
\frac{ (-1)^m}{m}\frac{\sin{(\frac{ 2\mu}{\hbar \om_+}\pi m)}}
{\sin{(\frac{\om_-}{\om_+}\pi m)}\,
\sinh{(\frac{2 k_B T}{\hbar \om_+}\pi^2 m)}} 
\label{4.26}
\\
\nonumber &+ &\left. \frac{1}{lpq}\sum_{k=1}^{\infty} 
\frac{(-1)^{(p+q)k}}{k\sinh{(\frac{\pi^2}{l}k)}}\left\lbrack
\frac{\mu}{k_BT}\cos{(\frac{\mu\pi k}{k_BTl})} 
-(\pi\coth{(\frac{\pi^2}{l}k)} + \frac{l}{\pi k})
\sin{(\frac{\mu\pi k}{k_BTl})} \right\rbrack \right\rbrack . 
\eqn
\end{itemize}

\subsection{The thermodynamical potential via symbol calculus}
Let us define the thermodynamical potential operator as 
${\cal O}=-\frac{1}{\be}\log{(1+\exp{[-\be(\check{\cal H}-\mu)]})}$.
We have: 
\bqn
\nonumber {\cal O}= \frac{1}{\pi^2}\int_{\BC^2}\, \hat{{\cal O}}(z_d,z_g) 
\mid z_d, z_g \rangle \langle z_d, z_g \mid \, d^2z_d \, d^2z_g, \\
\check{{\cal O}}(z_d, z_g)=\langle z_d, z_g \mid {\cal O} \mid z_d, z_g \rangle,
\eqn
We then perform the angular integrations and take its trace. We get an integral 
representation of the thermodynamical potential $\Omega=\mbox{Tr}{\cal O}$: 
\beq
\Om =-\frac{1}{\be}\int_0^{\infty}\int_0^{\infty}du_d\,du_g\, 
\hat{{\cal O}}(u_d,u_g) 
= -\frac{1}{\be }\int_0^{\infty}\int_0^{\infty}du_d\,du_g\,
\check{{\cal O}} (u_d, u_g).
\eeq
The problem turns out to evaluate the upper and lower symbols.
The lower one is shown here with the integral representation: 
\bqn
\nonumber\check{{\cal O}}(u_d,u_g)= 
-\frac{e^{-(u_d+u_g)}}{2\be}\int_{-\infty}^{+\infty} 
\frac{e^{-(ik+1)\frac{\be}{2}(\frac{\hbar \om}{2} - \mu)}}
{(\cosh{\frac{\pi}{2}k})(ik + 1)} 
\exp{\left(u_d e^{-(ik+1)\frac{\be}{2}\hbar \om_+}\right)} 
\exp{\left(u_g e^{-(ik+1)\frac{\be}{2}\hbar \om_-}\right)}\, dk
\\
\eqn 

\section{Average number of electrons and magnetic moment}
In this section, we will exploit the formulas (\ref{4.20})-(\ref{4.26}) to 
obtain the exact expressions of the average number of electrons and 
the magnetization. We will restrict ourselves to the more realistic case:
$\mu \leq \hbar\om/2 $.

The average number of electrons is easily derived by taking the derivative of
$-\Om$ with respect to $\mu$. It is found to be: 
\bqn
\nonumber \langle N_e\,\rangle &=& 
-\frac{1}{24}\left(\frac{\om_c}{\om_0}\right)^2 
+\frac{1}{2}\left\lbrack\left(\frac{\mu}{\hbar \om_0}\right)^2 
+\frac{\pi^2}{3}\left(\frac{k_BT}{\hbar\om_0}\right)^2
-\frac{1}{6}\right\rbrack \\
\nonumber & & +\frac{1}{4} \sum_{m=1}^{\infty} (-1)^m
\frac{e^{-\be \mu m}}{\sinh{(\frac{\be}{2}\hbar\om_+m)}\,
\sinh{(\frac{\be}{2}\hbar \om_-m)}} \\
\nonumber & &- \pi\sum_{m=1}^{\infty} (-1)^m 
\left\lbrack \frac{k_B T}{\hbar\om_-}
\frac{\cos{(\frac{2 \mu}{\hbar \om_-}\pi m)}}{\sin{(\frac{\om_+}{\om_-}\pi m)}\,
\sinh{(\frac{2k_B T}{\hbar\om_-}\pi^2 m)}} 
+\frac{k_B T}{\hbar \om_+ }\frac{\cos{(\frac{2\mu}{\hbar \om_+}\pi m)}}
{\sin{(\frac{\om_-}{\om_+}\pi m)}\,
\sinh{(\frac{2 k_B T}{\hbar \om_+}\pi^2 m)}} \right\rbrack \\
&\equiv& \langle N_e\, \rangle_{L} + \langle N_e\,\rangle_{01} 
+\langle N_e\, \rangle_{02} + \langle N_e\,\rangle_{\mbox{\scriptsize osc}}^- 
+\langle N_e\, \rangle_{\mbox{\scriptsize osc}}^+. \label{5.1}
\eqn

The magnetic moment is decomposed into four parts and is expressed in Bohr 
magneton units:
\bqn
\nonumber M &=& \chi_L H 
-2\mu_B\left(\frac{\pa\Om_{02}}{\pa\hbar\om_c}\right)_{\mu}  
-2\mu_B\left(\frac{\pa \Om_{\mbox{\scriptsize osc}}}{\pa\hbar\om_c}\right)_{\mu}
\\
&\equiv & 2\mu_B({\cal M}_L +{\cal M}_0+{\cal M}_{\mbox{\scriptsize osc}}^- +
 {\cal M}_{\mbox{\scriptsize osc}}^+),
\eqn 
where
\bqn
{\cal M}_L &=& \frac{-\mu}{12\hbar\om_0}\left( \frac{\om_c}{\om_0} \right)
        \equiv \frac{1}{2\mu_B} \chi_L H ,\\
{\cal M}_0 &=& \frac{1}{8\om}\sum_{m=1}^{\infty}
          (-1)^m e^{-\beta \mu m } \frac{
                 [ \om_+\coth{(\beta\hbar\om_+m/2)}
                  -\om_-\coth{(\beta\hbar\om_-m/2)}]}
                {\sinh{(\beta\hbar\om_+m/2)}\sinh{(\beta\hbar\om_-m/2)}},
\eqn
and, for the irrational case $\om_+/\om_- \not\in \Q$, 
\bqn
{\cal M}_{\mbox{\scriptsize osc}}^- &=& 
-\frac{k_BT}{\hbar\om} \sum_{m=1}^{\infty}
 \frac{(-1)^m \sin{(2\pi m\mu/(\hbar\om_-))}}
                  {\sin{(\pi m\om_+/\om_-)} \sinh{(2\pi^2mk_BT/(\hbar\om_-))}}
             \times \\ \nonumber
          & &\left[
 \frac{\pi\mu}{\hbar\om_-}\cot{\left(2\pi m\frac{\mu}{\hbar\om_-} \right)}
-\frac{\pi\om_+}{\om_-}\cot{\left(\pi m\frac{\om_+}{\om_-} \right)}
-\frac{\pi^2k_BT}{\hbar\om_-}\coth{\left(2\pi^2m\frac{k_BT}{\hbar\om_-}\right)}
             \right],\\
{\cal M}_{\mbox{\scriptsize osc}}^+ &=& 
\frac{k_BT}{\hbar\om} \sum_{m=1}^{\infty} 
              \frac{(-1)^m \sin{(2\pi m\mu/(\hbar\om_+))}}
                  {\sin{(\pi m\om_-/\om_+)}\sinh{(2\pi^2mk_BT/(\hbar\om_+))}}
              \times \\ \nonumber
          & &\left[
 \frac{\pi\mu}{\hbar\om_+}\cot{\left(2\pi m\frac{\mu}{\hbar\om_+} \right)}
-\frac{\pi\om_-}{\om_+}\cot{\left(\pi m\frac{\om_-}{\om_+} \right)}
-\frac{\pi^2k_BT}{\hbar\om_+}\coth{\left(2\pi^2m\frac{k_BT}{\hbar\om_+}\right)}
             \right]. 
\eqn 
We will not give the expressions of ${\cal M}^{\pm}_{\mbox{\scriptsize osc}}$
in the rational case because the magnetization is a continuous function of
$\om_c$ and its behavior can be fully understood from the irrational one. 

\subsection*{Discussion} 
The temperature scale is compared to the two natural modes $\om_{\pm}$ of the
system and draws three possible intrinsic regimes: high temperature regime
$k_B T > \hbar \om_+$, low temperature regime $k_B T < \hbar \om_-$, and 
intermediate temperature regime $\hbar \om_- < k_B T < \hbar \om_+$. 
Remember that we work in the large electron number region: $\mu > \hbar\om/2$.
\begin{enumerate}
\item {\bf High temperature regime}: $k_BT>\hbar\om_+>\hbar\om_-$.\\
This inequality implies the following constraint on the field: 
\beq
\frac{\om_c}{\om_0} < \frac{k_B T}{\hbar\om_0} - \frac{\hbar\om_0}{k_B T} 
\approx \frac{k_B T}{\hbar\om_0}.
\eeq
We can see that ${\cal M}_0$ is the dominant term for the magnetic moment in 
regard to ${\cal M}_{osc}$ because of the smallness of arguments of the sinh
(in the denominator) and coth (in the numerator) functions. 
Hence, $M \approx 2\mu_B ({\cal M}_L + {\cal M}_0)$, which shows mainly Landau
diamagnetism. 
Similarly, we infer from (\ref{5.1}) that 
$\langle N_e\,\rangle\approx\langle N_e\,\rangle_{L}+\langle N_e\,\rangle_{01}
+\langle N_e\, \rangle_{02}$. 
  
\item {\bf Low temperature regime}: $k_BT<\hbar\om_-$. \\
The magnetic field is restricted by the inequality:  
\beq
\frac{\om_c}{\om_0}<\frac{\hbar\om_0}{k_B T} -\frac{k_B T}{\hbar\om_0}  
\approx \frac{\hbar\om_0}{k_B T}.
\eeq
Now the ${\cal M}_0$ term becomes excessively small due to the rapidly decreasing
exponential factor and the large arguments of the hyperbolic functions present
in the expression. The magnetization is hence approximately
determined by the three terms: ${\cal M}_L$, ${\cal M}_{osc}^+$, and 
${\cal M}_{osc}^-$ and exhibits oscillating behavior. 
\begin{enumerate}
\item {\bf Strong fields} $\om_c \gg \om_0$.\\
We make the following approximations:
\beq
\om_+\approx\om_c\left(1+\left(\frac{\om_0}{\om_c}\right)^2\right),\ 
\om_-\approx\frac{\om_0^2}{\om_c}, \ 
\frac{\om_+}{\om_-}\approx\left(\frac{\om_c}{\om_0}\right)^2.
\eeq
Assume also that $\om_c \leq 2 \om_0 \sqrt{(\mu/\hbar)^2 - 1}$ to insure the 
validity of $\mu \geq \hbar \om / 2$. One can see that the denominator of 
${\cal M}_{osc}^+$ contains the product of sin and two sinh's with small 
argument and, hence, $M \approx 2\mu_B ( {\cal M}_L + {\cal M}_{osc}^+)$. 
In particular, after replacing $\om_+$ by $\om_c$ in the sinus argument of
the numerator of ${\cal M}_{osc}^+$, we find that the magnetization is 
periodic with respect to the inverse of the magnetic field, a characteristic 
fact of the ``de Haas-van Alphen'' regime. 
The similar behavior holds for the electron number:
$\langle N_e\, \rangle \approx \langle N_e\, \rangle_{L} +(\mu/\hbar\om_0)^2/2
+ \langle N_e\, \rangle_{\mbox{\scriptsize osc}}^+$. 
 
\item {\bf Weak fields} $\om_c \ll \om_0$.\\
In this case the two characteristic frequencies and their ratio are 
approximated to
\beq
\om_{\pm}\approx\om_0\left\lbrack 1\pm\frac{1}{2}\frac{\om_c}{\om_0}
+\frac{1}{8}\left(\frac{\om_c}{\om_0}\right)^2 \right\rbrack, \ 
\frac{\om_{\pm}}{\om_{\mp}} \approx 1 \pm \frac{\om_c}{\om_0} + 
\frac{1}{2}\left(\frac{\om_c}{\om_0}\right)^2.
\eeq
Now ${\cal M}_{osc}^-$ and ${\cal M}_{osc}^+$ have the same order of
contribution due to the presence in their denominators of 
\beq
(\sinh{(2\pi^2mk_BT/\hbar\om_{\mp})})^2
\approx(\sinh{(2\pi^2mk_BT/\hbar\om_0)})^2
\eeq
and of the strongly oscillating functions
\beq
(\sin{(\pi m\om_{\pm}/\om_{\mp})})^2\approx(\sin{(\pi m\om_c/\om_0)})^2.
\eeq 
Consequently, the system is considered as lying in the mesoscopic phase. 
Similar conclusions can be reached for the behaviour of the average electron 
number.  
\end{enumerate}

\item{\bf Intermediate temperatures}: $\hbar \om_- < k_B T < \hbar \om_+ $.\\
These inequalities imply the constraint 
\beq
\frac{\om_c}{\om_0} > \left\vert  \frac{\hbar\om_0}{k_B T} 
-\frac{k_B T}{\hbar\om_0} \right\vert . 
\eeq
The weak field case occurs only when $k_BT$ approaches $\hbar\om_0$.
In this case, one may think that the oscillatory terms ${\cal M}_{osc}^-$ and 
${\cal M}_{osc}^+$ give their contribution to $M$ as what we have seen in the 
previous subsection. But, in fact, the 
hyperbolic sinus functions of the denominators have large arguments:
$$
\sinh{(2\pi^2mk_BT/\hbar\om_{\mp})} 
\approx \sinh(19.74m)
\approx 0.5\,\exp(19.74m),
$$   
and so overcomes the algebraic contribution of the sinus functions. 
The system hence goes to the Landau diamagnetic regime. 
On the other hand, for a strong field, we return to the approximation 
$M \approx 2\mu_B ( {\cal M}_L + {\cal M}_{osc}^+)$ and the system shows
the de Haas-van Alphen effect.
\end{enumerate} 

\noindent
We have repeated in Fig.2 the phase diagram of magnetization proposed by Ishikawa 
and Fukuyama.  
Let us however mention that the two intrinsic frequencies $\om_{\pm}$ do not 
represent the exact borders between the different magnetic phases. It should be 
taken in a qualitative sense only.
In order to justify this, we choose the chemical potential equal to
$100.0\hbar\om_0$ and make  temperature vary through the different
magnetic field regimes. Fig.3(a) illustrates the weak field regime. 
One can see that at low temperature $k_BT = 0.001\hbar\om_0$ the magnetization
experiences strong fluctuations, called mesoscopic fluctuations. 
As $T$ increases ($k_BT=0.1\hbar\om_0$), the strength of these fluctuations decreases.
As the temperature is getting higher  (for example for $k_BT=0.5\hbar\om_0$),
the fluctuations disappear and we reach the Landau diamagnetism.
In Fig.3(b) the magnetic field lies between $1.9 \om_0mc/e$
and $3.1 \om_0mc/e$. 
When $k_BT=0.01\hbar\om_0$, the system shows large fluctuations. One can see 
that, if the magnetic filed increases, the fluctuations diminish and 
the cycloid-like curve appears (de Haas-van Alphen oscillations). 
At  higher temperature, {\it e.g.} $k_BT=0.1\hbar\om_0$, one can see more clearly 
the phase change from the mesoscopic 
fluctuations to the de Haas-van Alphen oscillations.
If the temperature continues increasing (for example, $k_BT=0.5\hbar\om_0$, 
$k_BT=1.0\hbar\om_0$), the system will go from the Landau diamagnetism to
the de Haas-van Alphen oscillations.
Fig.3(c) shows the de Haas-van Alphen phases. 
The position of the peak can be predicted from the simple formula: 
$$
\frac{\om_c}{\om_0}=\frac{2\mu}{n\hbar\om_0}-\frac{n\hbar\om_0}{2\mu},
\quad \mbox{where $n$ is some odd positive integer.}
$$ 
In Fig.3(d) we see that, at the extreme low temperature ($k_BT=0.001\hbar\om_0$),
the curve shows some ``width'', which 
comes from the limit of the picture resolution
and which corresponds,  in fact, to small fluctuations.
This gives us an example of  de Haas-van Alphen oscillations mixed
with  mesoscopic fluctuations.

\section{Current Distribution} 
We now turn our attention to the spatial density of current. 
It can be obtained by the following formula:
\beq
{\bf J}({\bf r})=\Re\left\langle\hat{\psi}^{\dagger}({\bf R})\frac{(-e)}{m}
\left({\bf P}+\frac{e}{c}{\bf A}({\bf R})\right) 
\hat{\psi}({\bf R}) \right\rangle,
\eeq 
where $\langle \cdot \rangle = \mbox{Tr}(f({\cal H}) (\cdot))$ is the thermal
average, and $\hat{\psi}({\bf R})$ the field operator: 
$$\hat{\psi}({\bf R}) \mid n, \alpha\, \rangle = \Psi_{n, \al}({\bf r}).$$
Due to the symmetry of the system, the current distribution is purely
orthoradial,
\beq
{\bf J}({\bf r}) = J_{\theta} (r) {\bf e}_{\theta},
\eeq
and its component is given by the series:
\beq
J_{\theta}(r)=-ev_0\sum_{n,\al}
\left(\al\frac{\xi}{r}+\frac{\om_c}{ 2\om_0}\frac{r}{\xi}\right) 
\bar{\Psi}_{n,\al}(r,\theta)\Psi_{n,\al}(r,\theta)f(E_{n \al}),
\label{6.3}
\eeq
where $\xi = \sqrt{\hbar/(m\om_0)}$ is the characteristic length and 
$v_0 = \om_0 \xi$ the characteristic speed in the harmonic potential. 
The current distribution induces a local magnetic moment,
\beq
{\bf M}({\bf r})=\frac{1}{2c} {\bf r} \times {\bf J} 
= \frac{1}{2c} r J_{\theta}(r){\bf e}_z
\equiv M_z (r) {\bf e}_z,
\eeq
which in turn gives rise to the total magnetic moment,
\beq
M=\int M_z (r)\,dS=\frac{\pi}{c}\int_0^{\infty}J_{\theta}(r)r^2\, dr.
\eeq 
By applying Eq.(\ref{3.8}) to the expression (\ref{6.3}) of $J_{\theta}(r)$ we get
\bqn
\nonumber J_{\theta}(r) &=& (-e)v_0\sum_{n_d,n_g}
f(\hbar(\om_+n_d+\om_- n_g+\frac{\om}{2})
\left((n_d-n_g)\left(\frac{r}{\xi}\right)^{-1}+
\frac{\om_c}{ 2\om_0}\frac{r}{\xi}\right) \times\\
\nonumber & & \frac{1}{\pi^4} \int\int\int\int_{\BC^2} 
e^{-(\vert z'_d\vert^2+\vert z'_g\vert^2+\vert z_d\vert^2+\vert z_g\vert^2 })
\bar{\cal K}(z'_d,z'_g;r,\theta){\cal K}(r,\theta;z_d,z_g)\times\\
& & \frac{(z'_d\bar{z}_d)^{n_d}}{n_d!}\frac{(z'_g\bar{z}_g)^{n_g}}{n_g!}
\,d^2 z'_d \, d^2 z'_g \, d^2 z_d \, d^2 z_g.
\eqn
The above equation is rotational invariant and one can hence put $\theta =0$.
Let us now represent the Fermi-Dirac function by the Fourier integral
(\ref{a.2}) and perform the discrete summation. The four integrals on the complex plane can be easily performed by
using well-known gaussian integrals of the type 
$$
\int_{-\infty}^{\infty}e^{-t^2 + tZ}\, dt =\sqrt{\pi}\exp{(\frac{1}{4} Z^2)}.
$$
We finally obtain the following expression: 
\bqn
\nonumber J_{\theta}(r)&=& \frac{-ev_0}{16\pi l_0^2}\frac{r}{\xi}
\int_{-\infty}^{\infty}\frac{dk}{\cosh{(\frac{\pi}{2} k)}} 
\frac{\exp{[(ik+1)\frac{\be}{2}\mu]}}{\sinh^2((ik+1)\be\frac{\hbar \om}{4})}
\times \\ \nonumber
& &\left\lbrack\frac{\om_c}{\om_0}\sinh{((ik+1)\be \frac{\hbar \om}{4})} -
\frac{\om}{\om_0}\sinh{((ik+1)\be \frac{\hbar\om_c}{4})}\right\rbrack \times \\
& & \exp\left[ \frac{-2r^2
\sinh((ik+1)\be\frac{\hbar\om_+}{4})
\sinh((ik+1)\be\frac{\hbar\om_-}{4})}
{l_0^2\sinh{((ik+1)\be\frac{\hbar\om}{4})} }\right].
\eqn
Three types of singular points appear in the integrand:
\begin{enumerate}
\item Pole $k=i$.\\
This pole is triple since it appears once in $\cosh{\frac{\pi}{2}k}$, and twice in
$(\sinh{(ik+1)\be \frac{\hbar \om}{4}})^2$.
\item Poles $k=(2m+1)i$, $m \in \Z^{\ast}$.\\
They are the other (simple) poles  of $\cosh{\frac{\pi}{2}k}$.
\item Singular points $k = i + 4 \pi m/(\be \hbar \om)$, $m \in \Z^{\ast}$.\\
These points are essential singularities since they appear in the exponential.
\end{enumerate}
The current distribution is hence separated into three components 
responsible for the Landau diamagnetism, weakly diamagnetism, and oscillation 
behavior:
\beq
\begin{array}{cccccccc}
J_{\theta}(r &=& &\left(J_{\theta}(r)\right)_L&+&\left(J_{\theta}(r)\right)_0&
+&\left(J_{\theta}r)\right)_{\mbox{\scriptsize osc}}\\
 &=& 2\pi i(&\overbrace{a_{-1}(i)}&+&\overbrace{\sum_{m\geq1}a_{-1}((2m+1)i)}&
+&\overbrace{\sum_{m \not= 0}(a_{-1}(i+\frac{4\pi}{\be\hbar\om} m)}).
\end{array}
\eeq
In a forthcoming paper, we shall go deeper into the analysis of such an  
expression for the current distribution.

\section{Conclusion} 
In this paper, we have established exact formulas for the thermodynamical potential of a two-dimensional 
gas of spinless electrons,
confined in an isotropic harmonic potential, and 
submitted to a constant perpendicular magnetic field. From this it has become possible to study the 
magnetic moment and other 
thermodynamical quantities at different 
regimes of temperature and field. This exhaustive study was made possible thanks to the specific 
simplicity of the isotropic 
harmonic potential. Of course, there exist other situations in which we could get similar exact 
expressions: anisotropic 
harmonic potential, harmonic groove, ..., as indicated by \cite{yofu}. In a next future, we shall 
deal with less tractable 
but still integrable models (see for instance \cite{rorile}, like infinite cylinder potential 
or quantum rings, cases in which the 
expressing of the trace (7.6)
in a closed form is quitely not expected. We shall also intend to make use of other families of 
coherent states, possibly more adapted to these new situations, in order to use an efficient symbol calculus.
For instance, for the infinite rectangular well potential, we are thinking about using tensor products 
of infinite well coherent states introduced and analysed in a recent paper \cite{agkp}. Finally, we shall also explore
the behaviour, at different temperature and field regimes, of other thermodynamical quantities of current 
experimental interest, like the  heat capacity \cite{mel}.

\section*{Appendix: Fermi-Dirac trace formulas} 
It is well known that, like the Gaussian function, the function 
$\mbox{sech}{x} = 1/\cosh{x}$ is a fixed point for the Fourier transform in 
the Schwartz space:
\beq
\frac{1}{\cosh{\sqrt{\frac{\pi}{2}}x}}
= \frac{1}{\sqrt{2\pi}} \int_{-\infty}^{+\infty} 
\frac{e^{-ixy}}{\cosh{\sqrt{\frac{\pi}{2}}y}}\, dy.
\eeq
Hence, given an Hamiltonian ${\cal  H}$, we can write for the corresponding Fermi operator:
\beq
f({\cal H})\equiv\frac{1}{1+e^{\be({\cal H}-\mu)}}=
\int_{-\infty}^{+\infty}\frac{e^{-(ik+1) \frac{\be}{2} ({\cal H}-\mu)}}
{4\cosh{\frac{\pi}{2}k}}\, dk. \label{a.2}
\eeq
Similarly, we can write for the thermodynamical potential operator:
\beq
-\frac{1}{\be}\log{(1+e^{-\be({\cal H} - \mu)})} = -\frac{1}{\be}\int_{-\infty}^{+\infty} \frac{e^{-(ik +
1)\frac{\be}{2}({\cal  H} - \mu)}}{(2\cosh{\frac{\pi}{2}k})(ik+1)}\, dk.
\eeq
Therefore, the average number of fermions and the thermodynamical potential can be written (at least formally) as follows:
\bqn
\langle N \rangle = \mbox{Tr}f({\cal  H}) 
=\int_{-\infty}^{+\infty}
\frac{e^{(ik+1)\frac{\be\mu}{2}}}{4\cosh{\frac{\pi}{2}k}}\Theta(k)\,dk,
\label{a.4} \\
\Om = \mbox{Tr}(-\frac{1}{\be}\log{(1+ e^{-\be({\cal  H} - \mu)})}) 
=-\frac{1}{\be}\int_{-\infty}^{+\infty}
\frac{e^{(ik+1)\frac{\be\mu}{2}}}{(2\cosh{\frac{\pi}{2}k})(ik+1)}\Theta(k)\,dk,
\label{a.5}
\eqn
where $\Theta$ designates the function 
\beq
\Theta (k) =\mbox{Tr} (e^{-(ik +
1)\frac{\be}{2}{\cal  H}}).
\label{a.6}
\eeq
Observe that  $(2m+1)i, \ m\in \Z$ are (simple) poles for the function 
$1/\cosh{\frac{\pi}{2}k}$ and $i$ is a pole for the functions $\Theta(k)$ 
and $1/(ik +1)$.
These Fourier integrals can be evaluated by using residue theorems  
if the integrand functions 
$\Phi_1(k)=\Theta (k)/\cosh{\frac{\pi}{2}k}$ and  
$\Phi_2(k)=\Theta (k)/((ik+1)\cosh{\frac{\pi}{2}k})$
satisfy the Jordan Lemma, that is,   
$\Phi_1(R e^{i\te}) \leq g(R)$, $\Phi_2(R e^{i\te})\leq h(R)$, for all 
$\te \in \lbrack 0, \pi \rbrack$, and 
$g(R)$ and $h(R)$ vanish as $R\to \infty$. The quantities
$\langle N \rangle$ and $\Om$ are then formally given by
\beq
2\pi i \left\lbrack a_{-1}(i)+\sum_{m=1}^{\infty}a_{-1}((2m+1)i) 
+\sum_{\nu} a_{-1}(k_{\nu})  \right\rbrack, 
\label{a.7}
\eeq
where $a_{-1}(\cdot)$ denotes the residue of the involved integrand at pole 
$(\cdot)$, and the $k_{\nu}$'s are the poles (with the exclusion of the pole $i$) of $\Theta(k)$ in the
complex $k$-plane.

We now introduce the spectral resolution of the (bounded below) self-adjoint 
operator ${\cal  H}$:
\beq
\varphi({\cal H})=\int_{-\infty}^{+\infty} \varphi(\lambda)\, E(d\lambda),
\eeq
where $\varphi$ is a complex-valued function and 
$E_{\lambda} = \int_{-\infty}^{\lambda}\, E(d\lambda)$ 
is the resolution of the identity for the Hamiltonian ${\cal  H}$. 
Define the density of states  $\nu(\lambda)$ as 
$\mbox{Tr}E(d\lambda)/d\lambda $. 
The trace formula ensues: 
\beq
\mbox{Tr}\varphi({\cal H})=
\int_{-\infty}^{+\infty}\varphi(\lambda)\,\nu(\lambda)\, d\lambda. \label{a.9}
\eeq
Let us now introduce the weighted density of states 
$ w(\lambda) = e^{-\frac{\be}{2} \lambda} \nu (\lambda)$ 
and its Fourier transform 
\beq
\hat{w} (k) = \frac{1}{\sqrt{2 \pi}}\int_{-\infty}^{+\infty} e^{-ik \lambda} w(\lambda)\, d\lambda.
\eeq
Then, from (\ref{a.4}), (\ref{a.5}) and (\ref{a.9}), we can represent
$\langle N \rangle$ and $\Om$ as follows: 
\bqn
\langle N \rangle =\sqrt{2\pi} \int_{-\infty}^{+\infty} 
\frac{e^{(ik + 1)\frac{\be \mu}{2}}}{4\cosh{\frac{\pi}{2}k}}
\hat{w}(\frac{\be }{2} k)\, dk 
=\frac{\pi}{\be} e^{\frac{\be \mu}{2}}\widehat{{\cal Z}}_1 (- \mu) ,\\
\Om = -\frac{\sqrt{2\pi}}{\be} \int_{-\infty}^{+\infty} 
\frac{e^{(ik + 1)\frac{\be\mu}{2}}}{(2\cosh{\frac{\pi}{2}k})(ik+1)} 
\hat{w}(\frac{\be }{2} k)\, dk 
= -\frac{2\pi}{\be^2} e^{\frac{\be \mu}{2}} \widehat{{\cal Z}}_2 (- \mu),
\eqn
where we have introduced the weighted functions 
${\cal Z}_1=\mbox{sech}(\frac{\pi}{\be} k) \hat{w} (k)$ and 
${\cal Z}_2 = \mbox{sech}(\pi k/\be) (i2k/\be+1)^{-1} \hat{w}(k)$.

\section*{Acknowledgements}
The authors are pleased to acknowledge 
Galliano Valent (LPTHE, Universities of Paris 6 and 7, France), 
Remi Mosseri (CNRS, GPS, Universities of Paris 6 and 7, France), 
Sorin Melinte (UPCPM, University of Louvain-la-Neuve, Belgium),  
and Yakov I. Granovskii (Szczecin University, Poland) for useful suggestions 
and comments.
P.Y.H. is also grateful the ICSC World Laboratory in Switzerland for the  
scholarship.

\clearpage
\section*{Figure caption} 
\begin{enumerate}
\item[Fig.1]
Poles of the Fourier representation of the thermodynamical potential $\Om$.
The poles lying on the imaginary axis are simple except for the point $i$ which is of order 4.  
The poles lying on the line $k=i$ in the complex $k$-plane ($i$ is excluded)
may be simple or double depending on whether $\om_+$ and $\om_-$ are
uncommensurable or not. 
\item[Fig.2]
Phase diagram of the magnetization.
In high temperature and low magnetic field region, the system shows  Landau
diamagnetism; in low temperature and low magnetic field region, 
mesoscopic fluctuations appear; and in strong magnetic field region, the system experiences
the de Haas-van Alphen oscillation phase.
The two curves, $k_BT=\hbar\om_+$ and $k_BT=\hbar\om_-$, give a qualitative indication
about the phase  borders. 
\item[Fig.3]
 Magnetization curves versus the magnetic field at different temperatures.
We represent the  cyclotron frequency in 
 $\om_0$ units and the magnetization in $2\mu_B$ units. 
The chemical potential $\mu$ is set up to $100.0\hbar\om_0$.
\begin{enumerate}
\item[(a)] $\om_c$ is less than $0.8\om_0$. Temperatures are chosen to be
    $0.001\hbar\om_0$, $0.1\hbar\om_0$, and $0.5\hbar\om_0$, respectively.
\item[(b)] $\om_c$ is between $1.9\om_0$ and  $3.1\om_0$. 
    Temperatures are chosen to
    be $0.01\hbar\om_0$, $0.1\hbar\om_0$, $0.5\hbar\om_0$, and $1.0\hbar\om_0$,
    respectively.
\item[(c)] $\om_c$ is between $4.0\om_0$ and  $15.0\om_0$. 
    Temperatures are chosen to
    be $0.1\hbar\om_0$, $0.5\hbar\om_0$, $1.0\hbar\om_0$, and $5.0\hbar\om_0$,
    respectively.
\item[(d)] $\om_c$ is greater than $15.0\om_0$. 
    Temperatures are chosen to
    be $0.001\hbar\om_0$, $0.5\hbar\om_0$, $1.0\hbar\om_0$, and $5.0\hbar\om_0$,
    respectively.
\end{enumerate}
\end{enumerate}

\clearpage
\begin{figure} 
\centering
\unitlength=1.0cm
\begin{picture}(10,10)
\put(4.87,1.37){$\times$}
\put(4.87,2.37){$\times$}
\put(4.87,3.37){$\times$}
\put(4.87,4.37){$\times$}
\put(4.87,6.37){$\times$}
\put(4.87,7.37){$\times$}
\put(4.87,8.37){$\times$}
 
\put(0,5){\vector(1,0){10}} \put(5,0.5){\vector(0,1){9}}
 
\put(1,5){\line(0,1){0.1}} \put(2,5){\line(0,1){0.1}}
\put(3,5){\line(0,1){0.1}} \put(4,5){\line(0,1){0.1}}
\put(5,5){\line(0,1){0.1}} \put(6,5){\line(0,1){0.1}}
\put(7,5){\line(0,1){0.1}} \put(8,5){\line(0,1){0.1}}
\put(9,5){\line(0,1){0.1}}
 
\put(5.3,1.375){$-7i$}\put(5.3,2.375){$-5i$}
\put(5.3,3.375){$-3i$}\put(5.3,4.375){$-i$}
\put(5.3,5.375){$i$}\put(5.3,6.375){$3i$}
\put(5.3,7.375){$5i$}\put(5.3,8.375){$7i$}
 
\put(0.8,4.5){$-8$}\put(2.8,4.5){$-4$}
\put(6.95,4.5){$4$}\put(8.95,4.5){$8$}
                                         
\put(0.64,5.5){\circle{0.1}}\put(1.51,5.5){\circle{0.1}}
\put(2.39,5.5){\circle{0.1}}\put(3.26,5.5){\circle{0.1}}
\put(4.13,5.5){\circle{0.1}}\put(5.87,5.5){\circle{0.1}}
\put(6.74,5.5){\circle{0.1}}\put(7.61,5.5){\circle{0.1}}
\put(8.49,5.5){\circle{0.1}}\put(9.36,5.5){\circle{0.1}}
 
\put(0.26,5.5){\circle*{0.1}}\put(1.44,5.5){\circle*{0.1}}
\put(2.63,5.5){\circle*{0.1}}\put(3.81,5.5){\circle*{0.1}}
\put(6.19,5.5){\circle*{0.1}}\put(7.37,5.5){\circle*{0.1}}
\put(8.56,5.5){\circle*{0.1}}\put(9.74,5.5){\circle*{0.1}}
 
\put(4.925,5.425){\rule{0.15cm}{0.15cm}}
 
\put(7,8){complex $k$-plane}
\end{picture}                       
\caption{FIG.1}
\end{figure} 

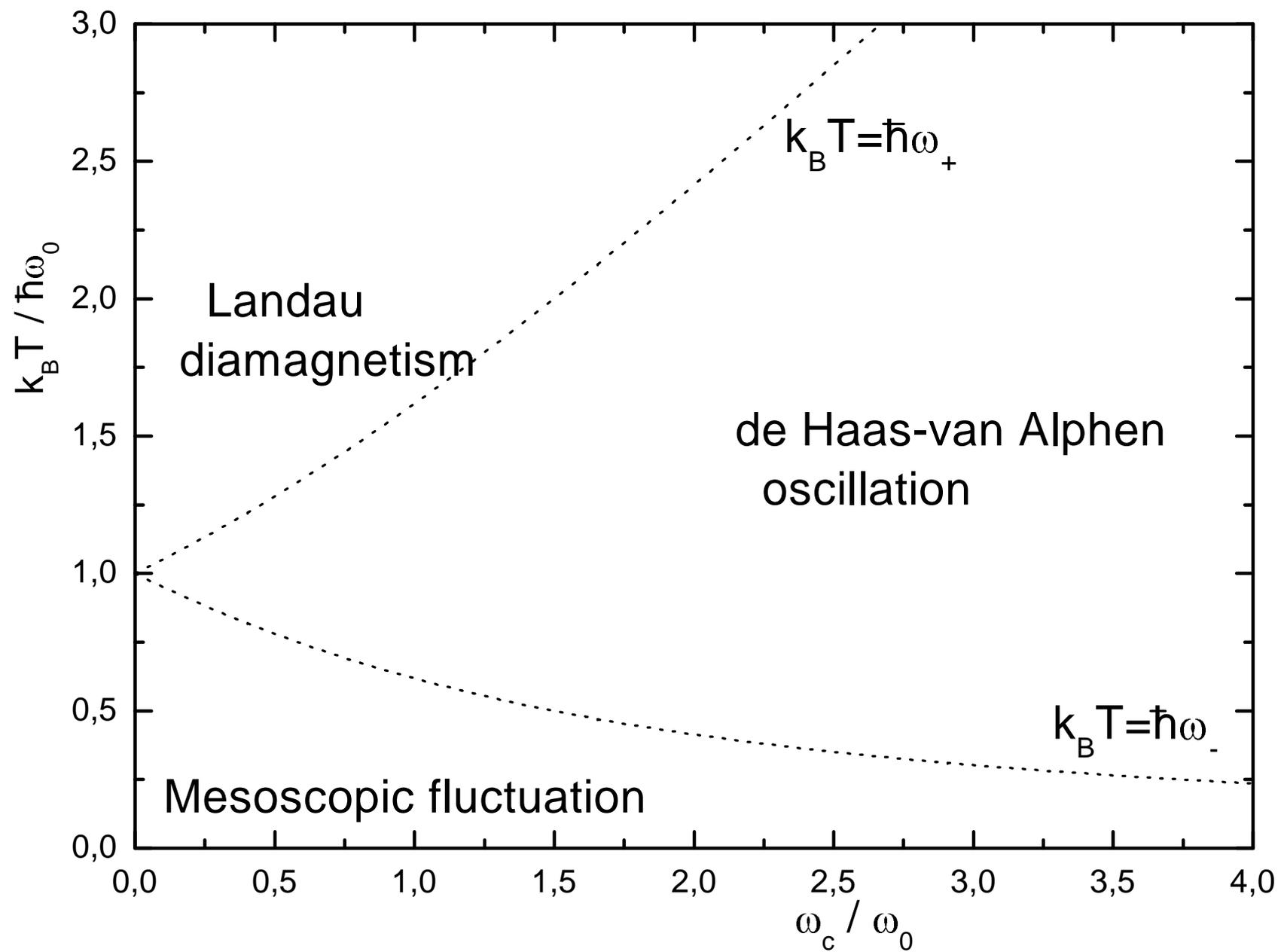
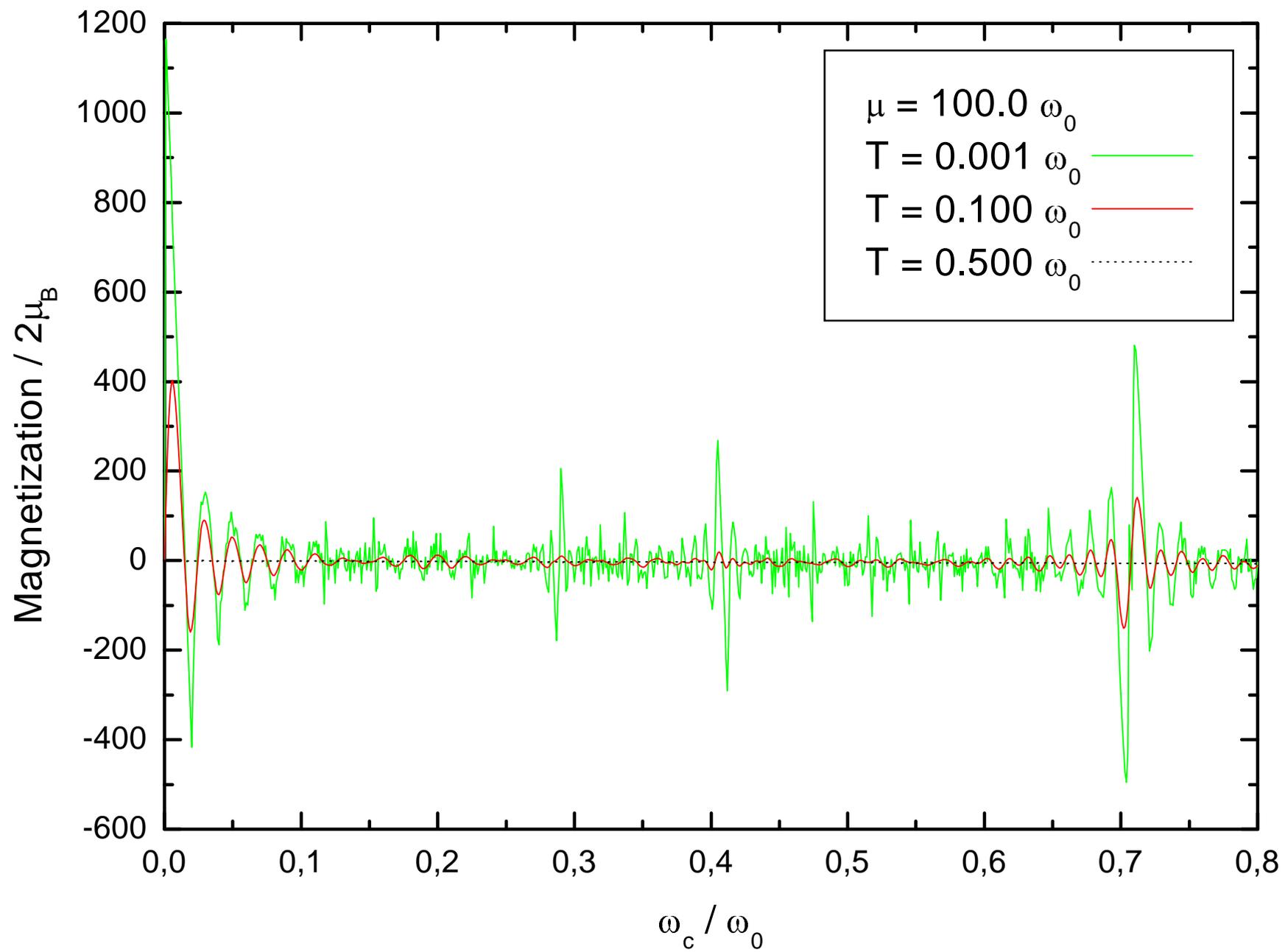
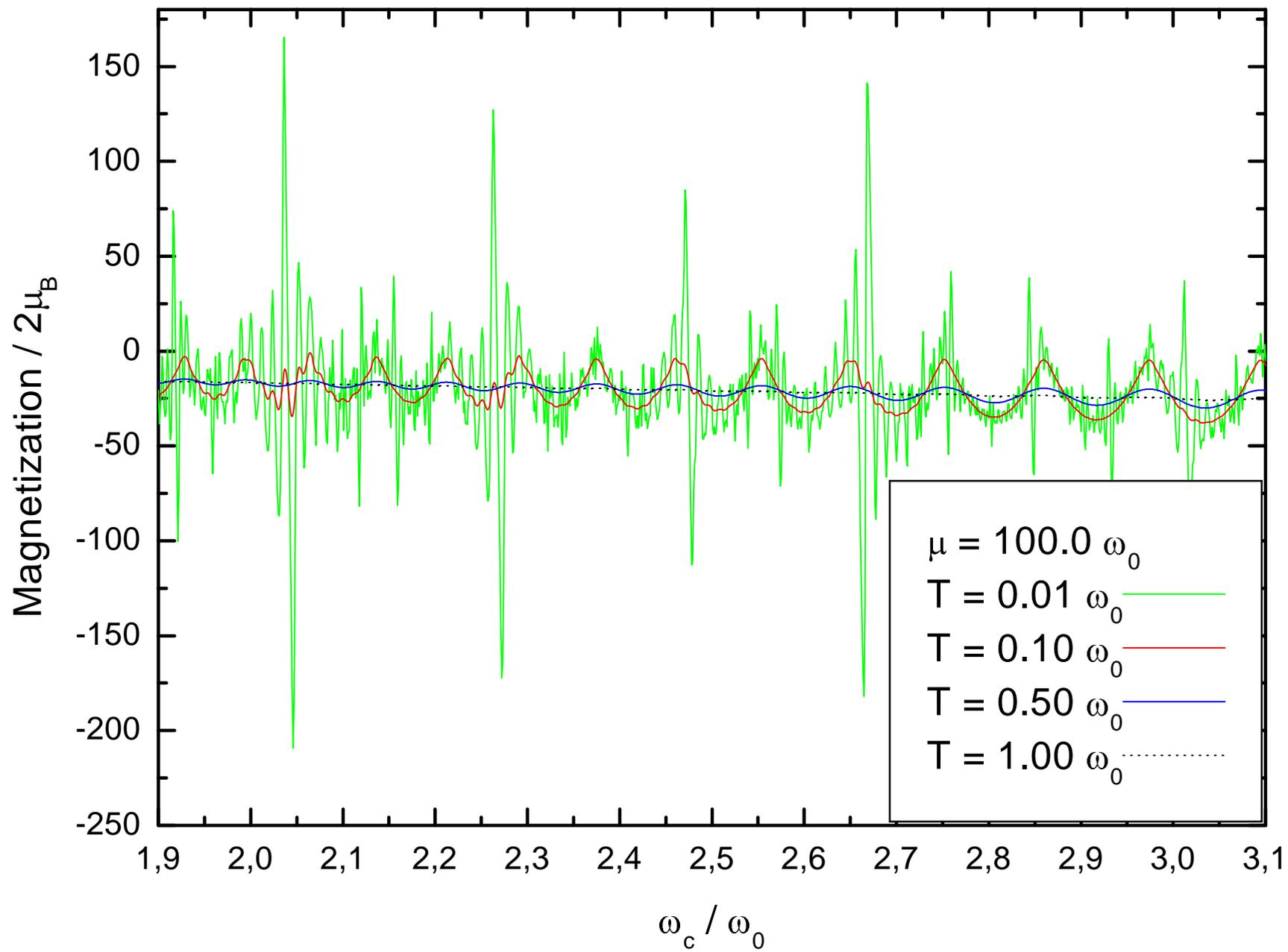
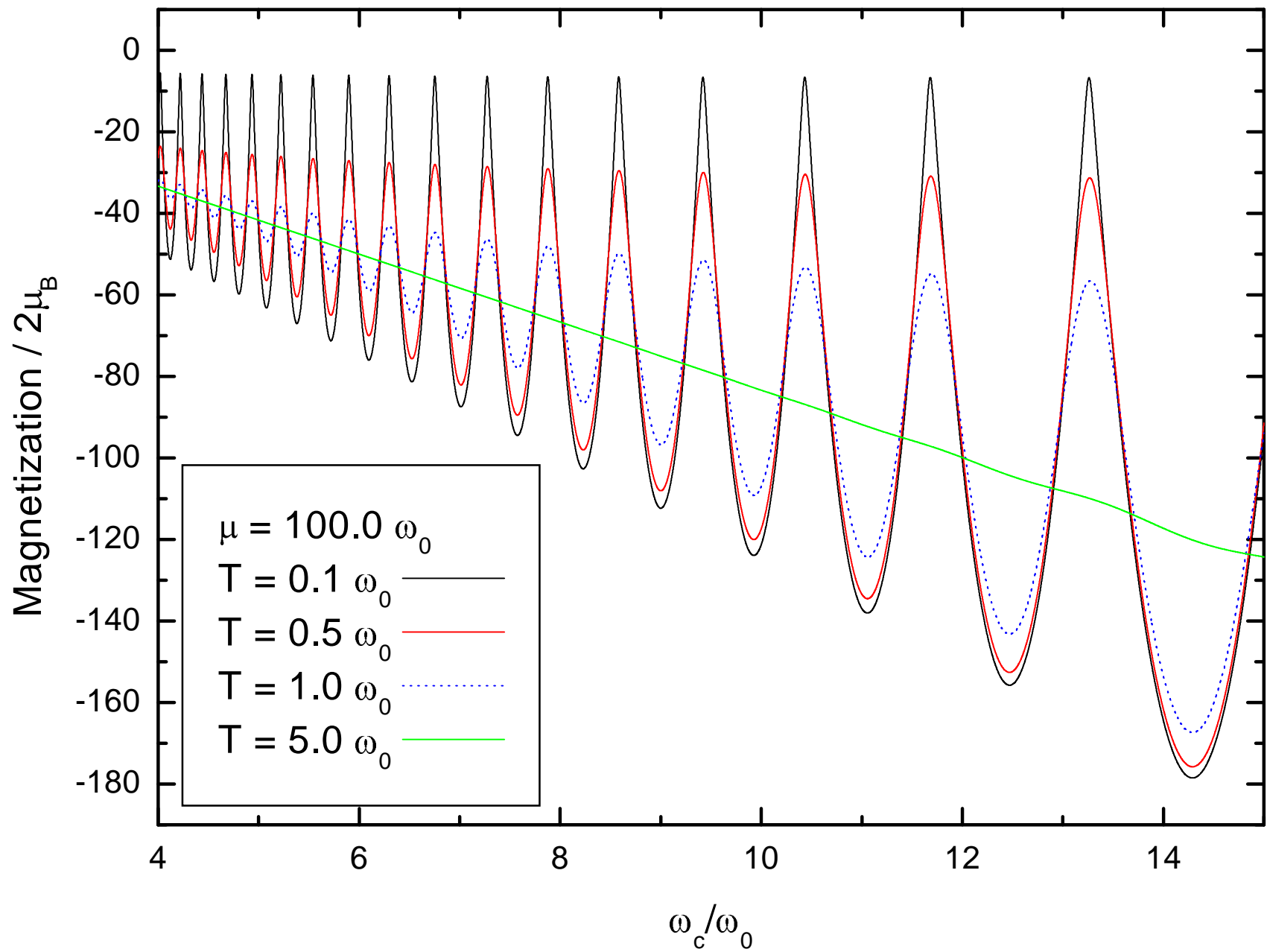
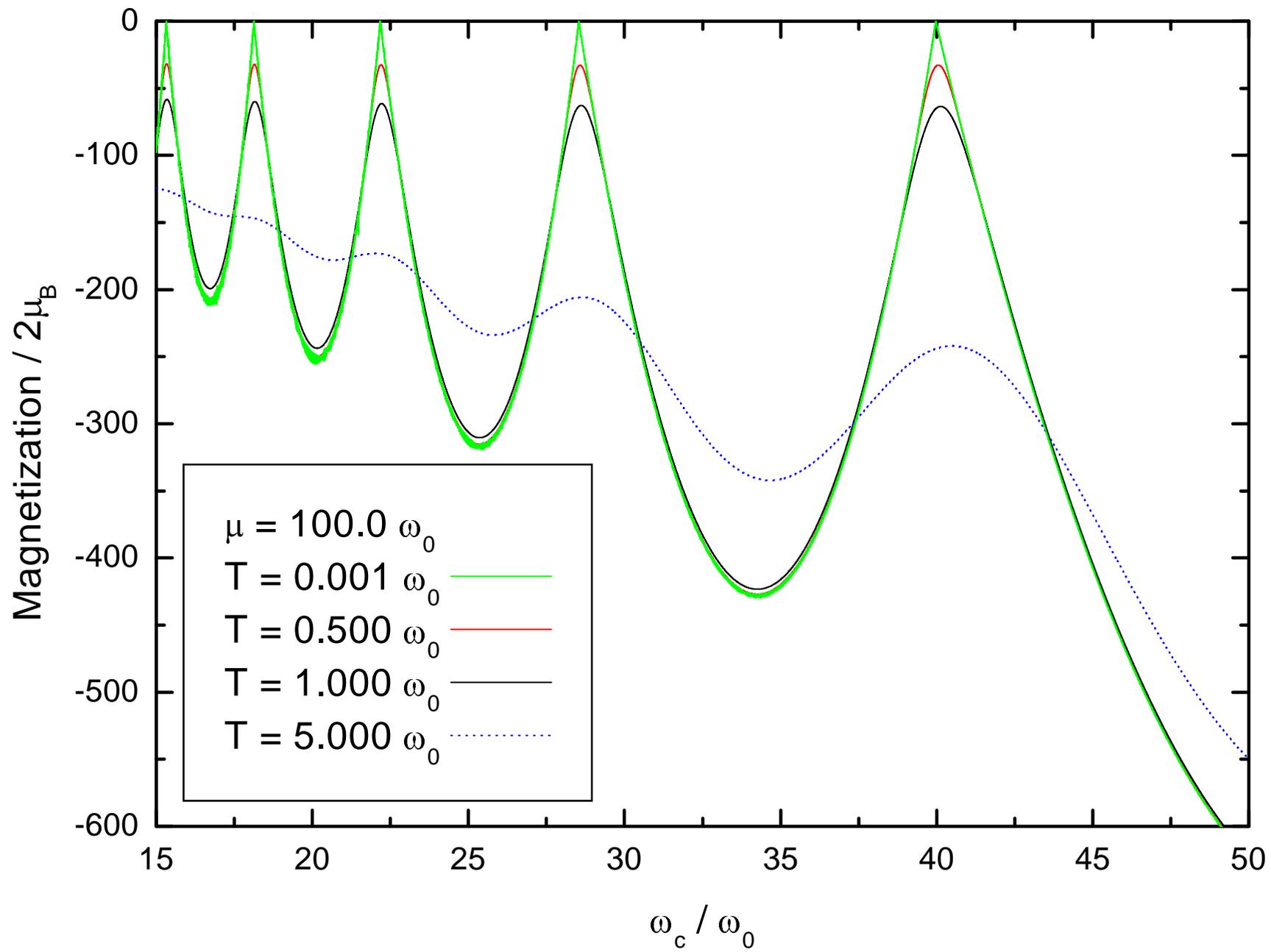
\end{document}